\documentclass[12pt]{article}
\usepackage{epsfig,latexsym}

\input epsf
\begin{document}
\font\cmss=cmss10 \font\cmsss=cmss10 at 7pt 
\hfill Bicocca-FT-01-15 \vskip .1in \hfill hep-th/0106160

\hfill

\vspace{20pt}

\begin{center}
{\Large \textbf{N=2 Gauge Theories}}
{\Large \textbf{from Wrapped Five-branes}}
\end{center}

\vspace{6pt}

\begin{center}
\textsl{F. Bigazzi, A. L. Cotrone and A. Zaffaroni} \vspace{20pt}

\textit{Universit\`{a} di Milano-Bicocca, Dipartimento di Fisica}

\textit{INFN - Sezione di Milano, Italy}
\end{center}

\vspace{12pt}

\begin{center}
\textbf{Abstract }
\end{center}

\vspace{4pt} {\small \noindent We present string duals
 of four dimensional $N=2$ pure SU(N)  SYM theory. The theory
is obtained as the low energy limit of D5-branes wrapped on non-trivial 
two-cycles. Using seven dimensional gauged supergravity and
uplifting the result to ten dimensions, we obtain solutions 
corresponding to various points
of the $N=2$ moduli space. The more symmetric solution may correspond
to a point with rotationally invariant classical vevs. By turning on 
seven dimensional scalar fields, we find a solution corresponding to
a linear distribution of vevs. 
Both solutions are conveniently studied with a D5-probe, which
also confirms many of the standard expectations for  $N=2$ 
solutions.}
\vfill     
\vskip 5.mm
 \hrule width 5.cm
\vskip 2.mm
{\small
\noindent francesco.bigazzi@mi.infn.it\\
aldo.cotrone@mib.infn.it\\
alberto.zaffaroni@mib.infn.it}

\section{Introduction}

Supergravity duals of non-conformal
N=2 gauge theories have been recently discussed in the literature from 
several complementary points of view. They can be obtained as mass deformations
of $N=4$ SYM \cite{pw,sfetsos,bpp,michela}, using fractional branes
at orbifold singularities \cite{kn,pg,bertolini,pproc,a,russo,billo}
or M5-branes wrapped on Riemann surfaces \cite{fayya}.
In this paper, we analyze the realization of pure SU(N) $N=2$
gauge theories using  wrapped type IIB NS-branes, an approach which
proved successful in the study of $N=1$ gauge theories \cite{mn1,mn2}.

Pure SU(N) $N=2$ SYM can be realized as the low energy theory
on D5-branes wrapped on
a non-trivial cycle of an ALE space\footnote{Another approach to
supergravity solutions for wrapped D5-branes can be found in
\cite{rajaraman}.}. We will also consider the
S-dual configuration with NS5-branes. 
Differently from $AdS_5\times S^5$
deformations and systems with fractional branes, at high energy
the $N=2$ theory is embedded in the six dimensional theory living
on the NS5. Such six dimensional theory decouples from gravity 
when the string coupling
is sent to zero, it is not conformal and not even
local \cite{seiberg}, but admit an holographic description in terms
of a linear dilaton background \cite{little}. 
We will find $N=2$ solutions of
the appropriate seven dimensional gauged supergravity 
which are asymptotic (in the ultraviolet)
to the linear dilaton background and we will uplift them 
to ten dimensions. 
The resulting solutions have a complex one dimensional
moduli space for the motion of a D5-probe, as expected for the dual
of the SU(N) $N=2$ SYM theory. We will find symmetric 
solutions corresponding to classical vevs
distributed in a rotationally invariant way. A particular solution in this
family may correspond to the strongly coupled theory with zero
(or smaller than the dynamically generated scale) classical vevs.
We will also find solutions corresponding to linear distributions 
of vevs. 
The moduli space structure
will be studied using a D5-probe, which, as usual, captures the 
quantum field theory one-loop result. 
The supergravity solutions we find are all singular.
In $N=2$ theories, an enhan\c con mechanism 
\cite{jpp} is usually invoked for resolving the singularity.
Some of the 
 features usually associated with the enhan\c con mechanism are 
at work here.    

In Section 2 we discuss our approach, which uses seven dimensional gauged 
supergravity. Since only the bosonic equations of motion of the
relevant theory are known \cite{cve1}, we will perform the singular
limit described in \cite{cve2} on the fermionic shifts of the
maximally gauged supergravity. Evidence
of the $N=2$ 
supersymmetry of the solution will be given by the probe analysis.
In Section 3 we will consider the uplifting to ten dimensions of
the most symmetric solution. By using a D5-probe, we will identify 
this solution as the point in moduli space
of $N=2$ SYM with zero classical vevs. 
In Section 4 we will turn on scalar fields
corresponding to the chiral operators parameterizing the
Coulomb branch. With
the probe analysis, we will identify this solution as corresponding 
to a linear distribution of vevs in the gauge theory.

While this work was being written, a paper \cite{martelli} appeared
where the solution in Section 2-3 was independently discussed.   

\section{Twisting the NS5 brane}
In order to obtain $N=2$ supersymmetry in four dimensions, we wrap N NS5
branes over an $S^{2}$ of an ALE space, and then  twist the normal bundle as in \cite{vafa}.
The ten dimensional spacetime is locally of the form ${\mathbf{R}}^{4}\times S^{2} \times{\mathbf{R}}^{2}\times\mathbf{R}^{2}$, where the first $\mathbf{R}^{2}$ is part of the ALE space and the second one is in the transverse flat space. These give
an $U(1)\times U(1)$ normal bundle. Working with seven dimensional gauged supergravity we can perform the twist by identifying the gauge fields in the theory
with the spin connection on the sphere as in \cite{mn1,mn2}. 
We thus choose the $U(1)\times U(1)$ truncation \cite{minasian} of the $SO(5)$ seven dimensional gauged supergravity \cite{pernici}.
As in \cite{mn1}, the right choice to retain $N=2$ supersymmetry is
to take one of the abelian gauge fields equal to the spin connection on the sphere, setting the other one to zero.

Our ansatz for the string frame metric in seven dimensions is: 
\begin{equation}
ds_{7}^2 = dx_{4}^2 + N[d\rho^2 + e^{2h}(d\theta^2 + \sin^{2}\theta d\varphi^2)],
\label{m1}
\end{equation}
which in the Einstein frame reads:
\begin{equation}
ds_{7}^2 = e^{2f}(dx_{4}^2 + d\rho^2) + e^{2g}(d\theta^2 + \sin^{2}\theta d\varphi^2),
\label{m2}
\end{equation}
with $f=-{2\over5}\Phi_{7},\, 
g= -{2\over5}\Phi_{7} + h$ ($\Phi_{7}$ is the seven dimensional dilaton). 
We also chose N=1 for simplicity.
We  look for solutions of the equations of motion of  $U(1)\times U(1)$ gauged supergravity that also preserve $ N =2$ supersymmetry. The theory at hand contains, apart from the metric, two scalars, two abelian gauge fields, a tree-form potential (which we take equal to zero in the following) and the corresponding fermions. 
The $U(1)\times U(1)$ truncation was used in \cite{mn1} to find $N=2$ M-theory solutions
interpolating between $AdS_{7}$ and $AdS_{5}$, corresponding to wrapped M5-branes.
In order to study NS5 branes in Type IIB, we need to perform a singular
limit in the theory, which reduces M theory to type II,
 as discussed in \cite{cve2}. The bosonic part of the
Lagrangian thus becomes:
\begin{equation}
2\kappa^{2}e^{-1}L = R +4m^2e^{(2\lambda_{1}+2\lambda_{2})} - 5\partial_{\mu}(\lambda_{1}+\lambda_{2})^2 - \partial_{\mu}(\lambda_{1}-\lambda_{2})^2 - e^{-4\lambda_{1}}{F_{\mu\nu}^{(1)}}^{2}- e^{-4\lambda_{2}}{F_{\mu\nu}^{(2)}}^{2}.
\end{equation}
While the maximally gauged supergravity (M theory compactified on $S^4$)
has $AdS$ vacua, corresponding to the $(2,0)$ CFT, the new theory
(type II on $S^3$) has only run-away vacua. There is instead a solution
corresponding to NS5 branes.
The same singular limit on the supersymmetry variations for the fermions gives ($k=2m$ is the gauge coupling)\footnote{It is possible to verify that the same ansatz gives, in the $N=1$ case, the results of \cite{mn2}.}:
\begin{eqnarray}
\delta\psi_{\mu}= [\nabla_{\mu}+ {k\over2}(A_{\mu}^{(1)}\Gamma^{12}+A_{\mu}^{(2)}\Gamma^{34}) + {1\over2}\gamma_{\mu}\gamma^{\nu}\partial_{\nu}(\lambda_{1} + \lambda_{2}) + \nonumber \\
+{1\over2}\gamma^{\nu}(e^{-2\lambda_{1}}F_{\mu\nu}^{(1)}\Gamma^{12}+ e^{-2\lambda_{2}}F_{\mu\nu}^{(2)}\Gamma^{34})]\epsilon , \nonumber \\
\delta{\lambda^{(1)}}= [{m\over4}e^{2\lambda_{1}}-{1\over4}\gamma^{\mu}\partial_{\mu}(3\lambda_{1}+2\lambda_{2})-{1\over8}\gamma^{\mu\nu}e^{-2\lambda_{1}}F_{\mu\nu}^{(1)}\Gamma^{12}]\epsilon , \nonumber \\
\delta {\lambda^{(2)}}= [{m\over4}e^{2\lambda_{2}}-{1\over4}\gamma^{\mu}\partial_{\mu}(2\lambda_{1}+3\lambda_{2})-{1\over8}\gamma^{\mu\nu}e^{-2\lambda_{2}}F_{\mu\nu}^{(2)}\Gamma^{34}]\epsilon .
\label{fermi}
\end{eqnarray} 
From (\ref{m2}) it follows that the non trivial components of the spin connection are:
\begin{equation}
\omega_{\hat{\alpha}}^{\alpha\rho}= f', \quad \omega_{\hat{\theta}}^{\theta\rho}=g'e^{g-f}, \quad \omega_{\hat{\varphi}}^{\varphi\rho}=g'e^{g-f}\sin\theta,  \quad \omega_{\hat{\varphi}}^{\varphi\theta}=\cos\theta,
\label{spinconn}
\end{equation}
where $\alpha=0,1,2,3$ labels the four dimensional coordinates in (\ref{m2}),
and the hats distinguish the curved coordinates from the flat ones. 
We impose the ansatz:
\begin{equation}
\gamma_{\rho}\epsilon = -\epsilon, \quad \gamma_{\varphi\theta}\epsilon=i\epsilon, \quad  \Gamma^{12}\epsilon= i\epsilon, \quad  \Gamma^{34}\epsilon= i\epsilon,\quad  \partial_{\alpha,\theta,\varphi}\epsilon = 0.
\end{equation}
As explained above, we take $A^{(1)}= -{1\over k}\cos\theta d\varphi, A^{(2)}=0$,
so that inserting (\ref{m2}) in (\ref{fermi}) we obtain (equating the variations (\ref{fermi}) to zero):
\begin{eqnarray}
f' &=& -({\lambda_{1}}' +{\lambda_{2}}'),\nonumber \\
g' &=& -({\lambda_{1}}' +{\lambda_{2}}') +{1\over k}e^{f-2g-2\lambda_{1}},\nonumber \\
3{\lambda_{2}}'+2{\lambda_{1}}' &=& -me^{f+2\lambda_{2}},\nonumber \\
3{\lambda_{1}}'+2{\lambda_{2}}' &=& -me^{f+2\lambda_{1}}+{1\over k}e^{f-2g-2\lambda_{1}},
\label{first}
\end{eqnarray}
whose solutions are:
\begin{eqnarray}
f &=& -(\lambda_{2}+\lambda_{1}) ,\nonumber \\
e^{2g-2f} &=& u , \nonumber \\
e^{\lambda_{2}-\lambda_{1}} &=& \sqrt{1-{1\over2u}+ {2Ke^{-2u}\over u}}, \nonumber \\
e^{\lambda_{2}+\lambda_{1}} &=& e^{-{2\over5}u}\left[1-{1\over2u}+ {2Ke^{-2u}\over u}\right]^{-{1\over10}} 
\label{7sol}
\end{eqnarray}
with:
\begin{equation}
{du\over d\rho}\equiv e^{\lambda_{2}-\lambda_{1}}.
\label{u}
\end{equation}
We have chosen the integration constants for $e^{-2g-2f}$ and $f$ in order that $u$ ranges in the interval $[0,\infty)$ and the seven dimensional dilaton has the canonical NS5 asymptotic behaviour (for $\rho\rightarrow\infty$)
$\phi\approx -\rho$. 
We also fixed to one the integration constant for $e^{\lambda_{2}+\lambda_{1}}$ which has no physical relevance, appearing as an overall
factor. We will
mainly consider the case $K\geq {1\over4}$ where $u\in [0,\infty)$.
For solutions with $K<{1\over4}$, $u$ can never reach zero. 

We explicitly checked that the
second order equations in \cite{cve1} 
are satisfied. We will have further evidence about the
supersymmetry of the solution in the next section, when we will find
a moduli space for D5 probes.
\section{Ten dimensional solution}
We refer to \cite{cve2} to lift the previous solution to ten dimensions. The string frame metric is ($ds_{7}^{2}$ is given by (\ref{m1}) with N=1):
\begin{equation}
ds^2= {ds_{7}}^{2} + {1\over m^2}e^{2\lambda_{2}+2\lambda_{1}}\Delta^{-1}\left[e^{-2\lambda_{1}}[d{\tilde\mu}_{1}^{2}+{\tilde\mu}_{1}^{2}(d\phi_{1}+ \cos\theta d\varphi)^2]+ e^{-2\lambda_{2}}[d{\tilde\mu}_{2}^{2}+{\tilde\mu}_{2}^{2}d\phi_{2}^{2}] \right]
\end{equation}
and the dilaton reads:
\begin{equation}
e^{2\Phi}= e^{6\lambda_{2}+6\lambda_{1}}\Delta^{-1},
\end{equation}
with
\begin{equation}
\Delta= e^{2\lambda_{1}}{\tilde\mu}_{1}^{2}+e^{2\lambda_{2}}{\tilde\mu}_{2}^{2}
\end{equation}
and $\phi_{1,2}, \tilde\mu_{1,2}=(\sin\theta^\prime,\cos\theta^\prime)$
being angular coordinates of the transverse three-sphere.

The solution incorporates also a potential six form whose field strength is given by:
\begin{eqnarray}
e^{-2\Phi}\ast F_{3} &=& 2me^{-5\lambda_{2}-5\lambda_{1}}\epsilon_{(7)}+{e^{-5\lambda_{2}-5\lambda_{1}}\over{2m}}\sum_{i=1}^{2}e^{-2\lambda_{i}}\ast_{(7)}de^{2\lambda_{i}}\wedge d(\tilde\mu_{i}^{2})\nonumber \\ 
&& -{e^{-3\lambda_{2}-3\lambda_{1}}\over 2m^3}e^{-4\lambda_{1}}d(\tilde\mu_{1}^{2})\wedge(d\phi_{1}+ \cos\theta d\varphi)\wedge \ast_{(7)}(\sin\theta d\theta\wedge d\varphi).
\label{eq1}
\end{eqnarray}
The D5 solution may be obtained from the one above by performing the S-duality transformations:
\begin{eqnarray}
\Phi_{D} &=& -\Phi, \nonumber \\
ds^{2}_{D} &=& e^{\Phi_{D}}ds^{2}_{NS}, \nonumber \\
dC_{6} &=& \ast F_{3}= e^{-2\Phi}\ast_{NS}F_{3}.
\label{eq2}
\end{eqnarray}
We can now examine the asymptotic behaviours of the metric and the dilaton making use of the explicit solutions (\ref{7sol}). 
In NS variables, we expect a solution UV asymptotic to the linear dilaton background \cite{little} with a size of $S^2$ that grows reflecting the coupling constant
running. Alternatively, in D5 variables,
when $u\rightarrow \infty$ ($\rho \rightarrow \infty$) we get:
\begin{eqnarray}
ds^{2}_{D}&\approx& e^{u}\left[ dx_{4}^2 + du^2 + u(d\theta^2 + \sin^{2}\theta d\varphi^2)+ \right. \nonumber \\
&& \left. +{1\over m^2}[d{\tilde\mu}_{1}^{2}+{\tilde\mu}_{1}^{2}(d\phi_{1}+ \cos\theta d\varphi)^2 d{\tilde\mu}_{2}^{2}+{\tilde\mu}_{2}^{2}d\phi_{2}^{2}] \right], \nonumber \\
e^{\Phi_{D}}&\approx& e^{u}.
\label{UV}
\end{eqnarray}
As expected the dilaton diverges and the $S^{2}$ blows up.

The details of the $u\rightarrow 0$ limit\footnote{This corresponds to the IR $\rho \rightarrow 0$ region if we take the integration constant in (\ref{u}) equal to zero.} drastically depends on the value of the integration constant $K$ in (\ref{7sol}). When $K>{1\over4}$ we find:
\begin{eqnarray}
ds^{2}_{D}&\approx& u^{-{1\over2}}|\tilde{\mu}_{2}| \left[ dx_{4}^2 + udu^2 + u(d\theta^2 + \sin^{2}\theta d\varphi^2)+ \right. \nonumber \\
&&\left. +{{\tilde\mu}_{2}^{2}\over m^2}[d{\tilde\mu}_{1}^{2}+{\tilde\mu}_{1}^{2}(d\phi_{1}+ \cos\theta d\varphi)^2 +
 u(d{\tilde\mu}_{2}^{2}+{\tilde\mu}_{2}^{2}d\phi_{2}^{2})] \right], \nonumber \\
e^{\Phi_{D}}&\approx& u^{-{1\over2}}|\tilde{\mu}_{2}|.
\label{IR}
\end{eqnarray}
The metric has a bad type singularity according to the criteria of \cite{mn1} and the dilaton is diverging, so we discard this possibility. When $K={1\over4}$ we find instead:
\begin{eqnarray}
ds^{2}_{D}&\approx& |\tilde{\mu}_{1}|\left[ dx_{4}^2 + {1\over u}du^2 + u(d\theta^2 + \sin^{2}\theta d\varphi^2)+ \right. \nonumber \\
&& \left.+ {1\over{|{\tilde\mu}_{1}|m^2}}[ud{\tilde\mu}_{1}^{2}+u{\tilde\mu}_{1}^{2}(d\phi_{1}+ \cos\theta d\varphi)^{2}+ d{\tilde\mu}_{2}^{2}+{\tilde\mu}_{2}^{2}d\phi_{2}^{2}] \right], \nonumber \\
 e^{\Phi_{D}}&\approx& |\tilde{\mu}_{1}|.
\label{IRft}
\end{eqnarray}
The singularity (which is located at $u=0$, or $\tilde\mu_{1}=0$) of this metric is milder, so that we can retain this solution as a dual of $N=2$ SYM.

We can explore the nature of the moduli space of the gauge theory using
a probe D5 brane wrapped on $S^{2}$, whose low energy effective action is (in units $2\pi\alpha'= \tau_{5}=1$):
\begin{equation}
S=-\int d^{6}\xi e^{-\Phi_{D}}\sqrt{-det(G+F)} + \int C_{6} +{1\over2} \int C_{2}\wedge F\wedge F ,
\label{azione}
\end{equation}
where
\begin{equation}
G_{\alpha\beta}=\partial_{\alpha}x^{M}\partial_{\beta}x^{N}g_{MN}
\end{equation}
is the induced metric on the worldvolume ($\alpha, \beta=0,1,...,5$ label the worldvolume coordinates, while $M,N=0,1,...,9$), and $F$ is the gauge field strength on the brane.

We  now perform our calculations in static gauge choosing
$\xi^{0}=x^{0}\equiv t, \xi^{i}=x^{i}, i=1,...,5 , x^{m}=x^{m}(t), m=6,...,9$ and taking the low velocity limit. From (\ref{eq1}) and (\ref{eq2}) it follows that, for $\theta^\prime=0$ :
\begin{equation}
\int C_{6}=4\pi V_{3}\int dt (ue^{2u}-{1\over2}e^{2u}+const),
\label{cisei}
\end{equation}
which does not depend on the angular coordinates. This term contributes to the effective potential for the probe. The other contribution comes 
from the Dirac-Born-Infeld part of the action. 
The BPS 
configurations correspond to having zero potential. If we search for solutions with unfixed radial coordinate $u$, the potential vanishes when $\tilde{\mu}_{1}\equiv\sin\theta^\prime=0$. In this case the low velocity limit of the DBI part of the action gives (taking $F=0$ on $S^{2}$):
\begin{equation}
S_{DBI}= -4\pi V_{3}\int dt u\left[g^{2}_{00}+{1\over2}g_{00}g_{mn}\dot{x}^{m}\dot{x}^{n}-{1\over4}F^{2}\right].
\label{dbi}
\end{equation}
The first contribution in the integral cancels exactly with $\int C_{6}$, after choosing the constant in (\ref{cisei}) equal to $2K$. This cancellation is independent of the particular
 choice of $K$, and can be proved using only the first order equations of
motion. The kinetic term in (\ref{dbi}) reads:
\begin{equation}
L_{kin}= -2\pi V_{3}u(e^{2u}{\dot{u}}^{2}+e^{2u}\dot{\phi}_{2}^{2}- {1\over2}F^2)
\end{equation}
which, introducing $r=e^{u}$, may be written as:
\begin{equation}
L_{kin}= -2\pi V_{3}\log r({\dot{r}}^{2}+ r^{2}\dot{\phi}_{2}^{2}- {1\over2}F^2).
\end{equation}
This gives a complex one-dimensional moduli space as expected for the $N=2$ gauge theory, which can be parameterized by the complex coordinate $z=re^{i\phi_{2}}$. We can explicitly write the holomorphic coupling after the calculation of the coefficient of $F\wedge F$ from the third term in (\ref{azione}), which equals $-2\pi\phi_{2}$.

After making explicit the dependence on the number N of D5 branes we find (apart from numerical factors) for the gauge kinetic term:
\begin{equation}
 {\cal I}m(\tau(z))F^2+{\cal R}e(\tau(z))F\tilde{F};\,\,\, \tau(z)={Ni\over \pi}\log {z\over\Lambda};\,\,\, \Lambda\sim{\sqrt N},
\label{lag}\end{equation}
while the scalar kinetic term reduces to:
\begin{equation}
{\cal I}m(\tau(z))\partial z\partial \bar{z}.
\end{equation}
We have thus found the structure for the moduli space expected for the $N=2$
supersymmetric four dimensional YM theory. The supergravity description captures correctly all the perturbative contributions to the coupling. 
Formula~(\ref{lag}) is compatible with points in moduli space where
all the classical vevs are zero or distributed in $U(1)_R$ invariant
configurations. $K$ may distinguish in between these cases.
Results in \cite{martelli} suggest that the radius of the
distribution is bigger than $\Lambda$ for $K<1/4$. This is compatible
with a probe able to move in the region inside the distribution
of branes, as found in \cite{martelli}.    
It is tempting to associate the fine tuned value $K=1/4$ with zero
(or smaller than $\Lambda$) classical vevs. 
For such a strongly coupled vacuum, an
enhan\c con mechanism \cite{jpp} could be expected: 
even if classically at the
origin, quantum mechanically the branes dispose in a spherical shell
of radius $\Lambda$. It is difficult to make this more precise because
the quantum field theory region of moduli space 
below the radius $\Lambda$ is hardly seen in the solution.
However, many features usually associated with the enhan\c con mechanism
are manifest in the solution with $K=1/4$:  at $u=0$ ($z=\Lambda$)
the probe become tensionless and extra bulk fields
become massless (D3-branes wrapped on the two-sphere, for example).
 The precise form of the singularity (and its resolution)
deserves further investigation. 
\section{A more interesting example}
We can also turn on other scalar fields in the seven dimensional gauged supergravity. The scalar fields parameterizing the sphere reduction are
expressed in terms of a symmetric $SO(4)$ tensor $T_{i,j}$, $i=1,...,4$.
In the previous Sections we retained the $U(1)\times U(1)$ 
singlets $T_{11}=T_{22}=e^{2\lambda_1}$ and 
$T_{33}=T_{44}=e^{2\lambda_2}$. The scalar fields $T_{ij},\, i,j=3,4$
parameterize the motion of the NS-branes in the untwisted $\mathbf{R}^2$ plane,
which preserves $N=2$ supersymmetry. They are dual to the bilinear scalar 
operators of the $N=2$ gauge theory. 
We expect that, as in similar $AdS_5$ examples \cite{freed2,pw}, we can
find solutions corresponding to non-trivial points in the Coulomb branch
of the $N=2$ theory. Up to a gauge rotation, we can take
$T_{11}=T_{22}=e^{2\lambda_1},T_{33}=e^{2\lambda_2},T_{44}=e^{2\tilde\lambda_2}$. This choice explicitly breaks $U(1)_{(2)}$. 
The equations of motion can be consistently truncated to
these three scalar fields. We look for $N=2$ solutions. We now need to
consider the full $SO(5)$ gauged supergravity and perform the
singular limit described in \cite{cve2}, in order to
descend from M theory to Type II. That this is a sensible
procedure will be guaranteed by the fact that the BPS first order
equations satisfy the second order equations \cite{cve1} for
type II compactifications. The BPS equations are now
\begin{eqnarray}
f' &=& -({\lambda_{1}}' +{{{\lambda_{2}}'+{\tilde\lambda_2}'}\over 2}),\nonumber \\
g' &=& -({\lambda_{1}}' +{{\lambda_{2}}'+{\tilde\lambda_2}'\over 2}) +{1\over k}e^{f-2g-2\lambda_{1}},\nonumber \\
{\lambda_{2}}'+2{\tilde\lambda_{2}}'+2{\lambda_{1}}' &=& -me^{f+2\tilde\lambda_{2}},\nonumber \\
{2\lambda_{2}}'+{\tilde\lambda_{2}}'+2{\lambda_{1}}' &=& -me^{f+2\lambda_{2}},\nonumber \\
3{\lambda_{1}}'+{\lambda_{2}}'+{\tilde\lambda_{2}}' &=& -me^{f+2\lambda_{1}}+{1\over k}e^{f-2g-2\lambda_{1}},
\label{first2}
\end{eqnarray}
which can be solved as in previous case. 
The relations $e^{2h}=u, du/d\rho=e^{(\lambda_{2}+\tilde\lambda_{2})/2- \lambda_{1}}$ still hold and moreover we have
the important relation
\begin{equation}
e^{\lambda_{2}-\tilde\lambda_{2}}={e^{2u}-b^2\over e^{2u}+b^2}.
\label{diff}
\end{equation}
For $b=0$ we recover the previously discussed solution with $\lambda_{2}=\tilde\lambda_{2}$. 

The up-lifting to ten dimensions can be performed using 
formulae in \cite{cve1}. The solution for wrapped D5
branes is:
\begin{eqnarray}
ds^2_{D}&=& \Delta^{1\over2}e^{-3\lambda_{1}-3(\lambda_{2}+\tilde\lambda_{2})/2}\left \{{ds_{7}}^{2} + {e^{2\lambda_{1}+\lambda_{2}+\tilde\lambda_{2}}\over m^2\Delta}\left [
e^{-2\lambda_{1}}[d\mu_1^2+d\mu_2^2+\cos^{2}\theta(\mu_{1}^2+\mu_{2}^2)d\varphi^{2}\right.\right. \nonumber \\
&& \left.\left. -2\cos\theta(\mu_{1}d\mu_{2}+\mu_{2}d\mu_{1})d\varphi] +e^{-2\lambda_{2}}d\mu_3^2 + e^{-2\tilde\lambda_{2}}d\mu_4^2
\right]\right \}
\end{eqnarray}
where $\sum_1^4 \mu_i^2=1$ parameterize $S^3$. The dilaton reads:
\begin{equation}
e^{-2\Phi_{D}}= e^{3\lambda_{2}+3\tilde\lambda_{2}+6\lambda_{1}}\Delta^{-1},
\end{equation}
with
\begin{equation}
\Delta= e^{2\lambda_{1}}(\mu_1^2+\mu_2^2) +
e^{2\lambda_{2}}\mu_3^2+e^{2\tilde\lambda_{2}}\mu_4^2.
\end{equation}
The complete formula for $F_{(3)}$ can be found in \cite{cve1}. We can choose $\mu_{1,2}=\sin\theta^\prime(\cos\phi_1,
\sin\phi_1)$ and $\mu_{3,4}=\cos\theta^\prime(\cos\phi_2,
\sin\phi_2)$.

The probe computation goes over as before. We find that,
for $\theta^\prime=0$,
\begin{equation}
V_{DBI}=C_{(6)}\sim\cos^2\phi_2e^{6f+2h+2\lambda_{2}}+
\sin^2\phi_2e^{6f+2h+2\tilde\lambda_{2}}
\label{pippo}
\end{equation}
so that there is a complex one-dimensional moduli space,
which can be parameterized as before by $z=re^{i\phi_2},
u=\log r$. In this Section we put, for simplicity, $\Lambda=1$.
The gauge field kinetic term is unchanged:
\begin{equation}
{\cal I}m(\tau(z))F^2+{\cal R}e(\tau(z))F\tilde{F};\,\,\, \tau(z)={Ni\over \pi}\log z ,
\label{gau}
\end{equation}
while the scalar kinetic term, using formulae~(\ref{first2}),~(\ref{diff}), reads:
\begin{eqnarray}
&&{\log|z|\over |z|^2}\left [ \cos^2\phi_2e^{-2(2\lambda_1+\lambda_2+2\tilde\lambda_{2})}+\sin^2\phi_2e^{-2(2\lambda_1+2\lambda_2+\tilde\lambda_2)}\right]dzd\bar z=\\\nonumber
&&\log|z|\left [ \cos^2\phi_2\left (1-{b^2\over r^2}\right )^2+\sin^2\phi_2\left (1+{b^2\over r^2}\right )^2\right ]dzd\bar z.
\label{sca}
\end{eqnarray}
In the standard coordinates for the $N=2$ effective Lagrangian, the scalar and gauge kinetic term coincide. We can 
obtain this by an holomorphic change of coordinates $w=z+
b^2/z$. The probe coupling constant now reads:
\begin{equation}
\tau(w)={Ni\over \pi}\left ( {\rm arcosh}({w\over 2b}) + {\mbox const}\right ).
\label{tau}
\end{equation}
For large $|w|$, $\tau(w)$  has the standard logarithmic behaviour, while for small $|w|$ it reveals a non-trivial
distribution of vevs in the gauge theory. We can explicitly compute such distribution, by approximating it 
with a continuum:
\begin{equation}
\tau(w)={i\over\pi}\sum_i \log(w-a_i)\sim i\int_{-2b}^{2b}da\mu (a)\log (w-a),
\label{tau2}
\end{equation}
with $\mu (a)=N/(\pi\sqrt{4b^2-a^2})$. We see that our solution
represents a point in the Coulomb branch where the vevs are
 linearly distributed. This is somehow
reminiscent of \cite{bpp}
with the obvious difference that the theory in \cite{bpp}
is a mass deformation of $N=4$,  while our starting point is (a little string UV completion of) pure $N=2$ without matter.
Curiously, this linear density of vevs is of the same type which
appears for the $N=2$ point in moduli space where all types of
monopoles become massless \cite{ds}. The relation of our 
solution with such a particular point and with the possible
soft breaking to $N=1$ deserves further investigation.

We gave several indications that our solutions are actually $N=2$.
It would be interesting to check the supersymmetry directly
in ten dimensions. We understand that this was done in \cite{martelli} 
for the solution in Section 2 and 3.

\vskip .1in
\noindent \textbf{Acknowledgments}\vskip .1in \noindent  
The authors are partially supported by INFN, MURST and the European Commission TMR program HPRN-CT-2000-00131, 
wherein they are associated to the University of Padova.


\begin{thebibliography}{99}
\bibitem{pw} K. Pilch and N. P. Warner, 
Nucl. Phys. B594 (2001) 209; hep-th/0004063. 
\bibitem{sfetsos} A. Brandhuber and K. Sfetsos,
Phys. Lett. B488 (2000) 373; hep-th/0004148. 
\bibitem{bpp} A. Buchel, A. W. Peet and J. Polchinski, 
Phys. Rev. D63 (2001) 044009; hep-th/0008076. 
\bibitem{michela} N. Evans, C. V. Johnson and
M. Petrini, 
JHEP 0010 (2000) 022; hep-th/0008081. 
\bibitem{kn}I. R. Klebanov, N. A. Nekrasov, Nucl. Phys. B574 (2000) 263; hep-th/9911096.
\bibitem{pg}M. Grana and J. Polchinski, 
Phys. Rev. D63 (2001) 026001; hep-th/0009211; hep-th/0106014. 
\bibitem{bertolini} M. Bertolini, P. Di Vecchia, 
M. Frau, A. Lerda, R. Marotta and I. Pesando,
JHEP 0102 (2001) 014; hep-th/0011077. 
\bibitem{pproc}J. Polchinski, 
Int. J. Mod. Phys. A16 (2001) 707; hep-th/0011193. 
\bibitem{a}O. Aharony, 
JHEP 0103 (2001) 012; hep-th/0101013. 
\bibitem{russo}M. Petrini, R. Russo and A. Zaffaroni, 
hep-th/0104026.
\bibitem{billo}M. Billo, L. Gallot and A. Liccardo, 
hep-th/0105258.
\bibitem{fayya} A. Fayyazuddin and D. J. Smith,
JHEP 0010 (2000) 023; hep-th/0006060; B. Brinne, A. Fayyazuddin, S. Mukhopadhyay and D. J. Smith, 
JHEP 0012 (2000) 013; hep-th/0009047. 
\bibitem{mn1} J. M. Maldacena and C. Nu$\tilde{\rm n}$ez,
Int. J. Mod. Phys. A16 (2001) 822-855;  hep-th/0007018.
\bibitem{mn2}J. M. Maldacena and C. Nu$\tilde{\rm n}$ez, 
Phys. Rev. Lett. 86 (2001) 588-591; hep-th/0008001.
\bibitem{rajaraman}A. Rajaraman, 
hep-th/0011279.
\bibitem{seiberg} M. Berkooz, M. Rozali and N. Seiberg, 
Phys. Lett. B408 (1997) 105; hep-th/9704089.
\bibitem{little}O. Aharony, M. Berkooz, D. Kutasov
and N. Seiberg, 
JHEP 9810 (1998) 004; hep-th/9808149. 
\bibitem{jpp}C. V. Johnson, A. W. Peet and J. 
Polchinski,
Phys. Rev. D61 (200) 086001; hep-th/9911161. 
\bibitem{cve1} M. Cvetic, H. Lu, C.N. Pope, 
Phys. Rev. D62 (2000) 064028; hep-th/0003286.
\bibitem{cve2} M. Cvetic, J. T. Liu, H. Lu, C.N. Pope, 
Nucl. Phys. B560 (1999) 230-256; hep-th/9905096. 
\bibitem{martelli} J. P. Gauntlett, N. Kim, D. Martelli
and D. Waldram, hep-th/0106117.
\bibitem{vafa} M. Bershadsky, C. Vafa and V. Sadov, 
Nucl. Phys. B463 (1996) 420; hep-th/9511222.
\bibitem{minasian}J. T. Liu, R. Minasian, 
Phys. Lett. B457 (1999) 39-46; hep-th/9903269.
\bibitem{pernici} M. Pernici, K. Pilch, P. van Nieuwenhuizen, 
Phys. Lett. B143 (1984) 103.
\bibitem{freed2} D. Z. Freedman, S. S. Gubser, K. Pilch
and  N. P. Warner, 
JHEP 0007 (2000) 038; hep-th/9906194.
\bibitem{ds} M. R. Douglas and S. H. Shenker, 
Nucl. Phys. B447 (1995); hep-th/9503163.
\end{thebibliography}
\end{document}